\documentstyle[psfig,11pt]{article}

\begin{document}

\title{Spherical Curvature Inhomogeneities in String Cosmology}
\author{John D. Barrow and Kerstin E. Kunze \\
Astronomy Centre\\
University of Sussex, Brighton BN1 9QJ\\
United Kingdom}
\date{}
\maketitle

\begin{abstract}
We study the evolution of non-linear spherically symmetric inhomogeneities
in string cosmology. Friedmann solutions of different spatial curvature are
matched to produce solutions which describe the evolution of non-linear
density and curvature inhomogeneities. The evolution of bound and unbound
inhomogeneities are studied. The problem of primordial black hole formation
is discussed in the string cosmological context and the pattern of evolution
is determined in the pre- and post-big-bang phases of evolution.


\end{abstract}

\section{Introduction}

Considerable attention has been devoted to exploring the range of behaviours
displayed by the equations of string cosmology which are supplied by the
variation of the low-energy effective action of the bosonic sector of string
theory \cite{refA}. Investigations have been made into the evolution of
isotropic cosmologies \cite{ref5}, simple homogeneous anisotropic
cosmologies of Bianchi type, \cite{batak}, and Kantowski-Sachs type \cite
{ref10}, by various authors, and the present authors have provided a
systematic classification of spatially homogeneous string cosmologies in
terms of their relative generality when considered as constrained systems of
non-linear ordinary differential equations\cite{ref11}.

As in the case of general-relativistic cosmologies, the introduction of
inhomogeneities into the string cosmological equations produces a
considerable increase in mathematical difficulty: non-linear partial
differential equations must now be solved. In practice, this means that we
must proceed either by means of approximations which render the
non-linearities tractable, or we must introduce particular symmetries into
the metric of space-time in order to reduce the number of degrees of freedom
which the inhomogeneities can exploit. Accordingly, inhomogeneous string
cosmologies have been investigated in the approximation of small
perturbations of the isotropic Friedmann-Robertson-Walker (FRW) models, in
the 'velocity-dominated' approximation, and by studies of exact
inhomogeneous solutions with cylindrical symmetry. Barrow and Kunze \cite
{ref9} found a wide class of exact cylindrically symmetric flat and open
inhomogeneous universes. Closed cylindrical solutions were then found by
Feinstein et al. \cite{ref8}. These solutions provide exact
'velocity-dominated' solutions of general relativity and are expected to
form a leading-order approximation to part of the general solution of string
cosmology in the neighbourhood of the singularity (if we ignore higher-order
string corrections to the action). Veneziano \cite{venez} and Buonanno {\it %
et al} \cite{ref3b} have studied the behaviour of string cosmologies at
early times in the velocity-dominated approximation, (that is, neglecting
spatial derivatives, relativistic motions, and 3-curvature inhomogeneities
with respect to time-derivatives in the field equations \cite{veldom}). The
asymptotic forms obtained at early times also approximate the behaviour
displayed by the exact solutions of Barrow and Kunze on scales larger than
the horizon where the inhomogeneities evolve slowly: the inhomogeneities are
just homogeneously propagated. However, the exact solutions also provide
information about the evolution of dilaton, axion, and gravitational wave
inhomogeneities after they enter the horizon, where they attenuate by
non-linear oscillations because of the pressure forces exerted by the
dilaton and axion fields. These solutions do not contain trapped surfaces
and so they cannot be used to follow the collapse of inhomogeneities to black
holes, although it would be possible to study this problem by using the
closed ($S^3$) solutions studied by Feinstein {\it et al }in \cite{ref8}.

The exact solutions given in refs. \cite{ref9} and \cite{ref8} possess
cylindrical symmetry and all physical quantities depend on at most one space
coordinate and the time. The case of cylindrical symmetry is natural because
of the mathematical simplicity of the field equations whenever there exists
a direction in which the pressure equals the energy density. However, it is
also important to consider the case where the inhomogeneities possess
spherical symmetry. Not only does this seem more natural, in that there need
exist no preferred direction in which the inhomogeneity dominates, but it
allows the problem of bound inhomogeneities to be addressed more directly
without the complication of gravitational wave inhomogeneities.

The choice of spherically symmetric inhomogeneity does not permit exact
solutions of Einstein's equations except where fluids have vanishing
pressure. However, a clear physical picture of the behaviour of spherically
symmetric inhomogeneities with non-zero pressure can be obtained by the
device of matching together homogeneous solutions of different curvature and
density. The resulting patched solution describes the evolution of spherical
overdensities (or underdensities) in a smooth background universe. In the
limit that the inhomogeneities become small, they will evolve in accord with
the results of small perturbation theory. When the inhomogeneities are not
small, we obtain a description of non-linear processes like void formation,
the condensation of gravitationally bound lumps, or the creation of
primordial black holes. This technique was first introduced into general
relativity to study the evolution of inhomogeneities by Lema\^itre \cite
{lemait} and was subsequently applied to the study of protogalaxies by
Harrison \cite{harr}. Here, we shall apply it to the equations of string
cosmology to further our understanding of the evolution of the pre- and post-
big-bang phases in the presence of spherical inhomogeneities.

The string cosmological models considered here are derived from the bosonic
sector of heterotic string theory reduced to (3+1) dimensions of spacetime.
They are assumed to have vanishing cosmological constant and vanishing
Maxwell field. Their field content consists of an antisymmetric tensor
field, a dilaton, and the space-time metric tensor. However, in the
low-energy limit only the antisymmetric tensor field strength is important
in the equations of motion.

In 10-dimensional superstring theory, gravitational anomalies occur which
signal the breakdown of energy-momentum conservation. However, by redefining
the antisymmetric tensor field strength, these anomalies can be cancelled.
The redefined antisymmetric tensor field strength, $H,$ is given by \cite
{ref1} 
\begin{eqnarray}
H=dB+\omega _L,  \nonumber
\end{eqnarray}
where $B$ is the antisymmetric tensor field and $\omega _L$ is the
Lorentz-Chern-Simons (LCS) form involving the Lorentz spin connection. The
antisymmetric tensor field strength is a 3-form. In four dimensions it is
dual (in the sense of differential forms) to a one-form which can be shown
to be the gradient of a scalar field, the axion. FRW spaces are maximally
symmetric and so a theorem proved in \cite{ref2} implies that all
contributions from the LCS terms vanish. Hence, the energy-momentum tensor
for the axion and the dilaton in the Einstein frame consists of two coupled
stiff perfect fluids.

The low-energy limit of string theory provides a new picture for the
evolution of the early universe. Once a stage of low coupling and small
curvature is reached, the universe enters the 'pre-big-bang' era \cite{ref1a}%
. During this stage the universe undergoes superinflation (accelerated
expansion) in the string frame driven by the kinetic energy of the dilaton.
By contrast, in the Einstein frame this corresponds to an accelerated
contraction. The pre-big-bang era ends when the string coupling becomes
strong enough for the low-energy limit to be no longer valid. However,
exactly how this ``graceful exit'' can be effected is not yet fully
understood \cite{ref2a}. In the transition era, complicated non-perturbative
effects will become important and it is not clear if a curvature singularity
can always be prevented by higher-order contributions. Eventually, the
universe must enter the (classical) post-big-bang era. Therefore, the
pre-big-bang phase can be understood as a way to provide initial conditions
for the classical post-big-bang era. An interesting aspect of this scenario
is provided by the duality symmetries present in string theory. In spatially
homogeneous cosmological models, scale-factor duality relates solutions for
the pre-big-bang phase (``+ branch'') to those for the post-big bang phase
(``-- branch''). Unless there is a self-dual solution there is the problem
of how to relate these two branches. This may require an explicitly quantum
cosmological transition \cite{quant}. In the presence of inhomogeneities,
especially those which allow some parts of the Universe to expand whilst
other parts collapse, the impact of duality invariance may prove more
unusual and motivates further detailed study of realistic inhomogeneous
string cosmologies.

Here, we extend our understanding of inhomogeneous string cosmologies by
investigating the simple model of non-linear spherically symmetric
inhomogeneities outlined above, in which a spherical curvature perturbation
is self-modelled by an FRW universe of non-zero curvature in a flat
background FRW universe. In section 2 we describe the self-modelling of
spherical inhomogeneities in general relativistic universes containing
fluids with pressure equal to density. In section 3 the connection with
string theory is displayed. In sections 4 and 5 the post and pre-big-bang
solutions are given and, finally, the results are discussed in section 6.

\section{Spherical Inhomogeneities in General Relativity}

Spherically symmetric density inhomogeneities can be modelled by matching a
section of a closed (or open) FRW universe to a background universe
described by a flat FRW universe in such a way that the metric and its
derivatives are continuous at the boundary. First, consider this matching in
general relativity for the simple case of a universe containing a perfect
fluid, with pressure $p$ and energy density $\rho ,\ $which are related by a
stiff equation of state, $p=\rho $.

The flat background FRW universe has an expansion scale factor $R(t)$ and
its dynamics are determined by the Friedmann equation 
\begin{eqnarray}
\left( \frac{dR}{dt}\right) ^2=\frac{8\pi G}3\rho R^2  \label{e1}
\end{eqnarray}
where $t$ is proper time. Note that Einstein's equations are invariant under
time-reversal, i.e. $R(t)=R(-t)$.

Energy-momentum conservation for the $p=\rho $ fluid implies that 
\begin{eqnarray}
\rho \propto R^{-6}.  \label{e2}
\end{eqnarray}
and so $R(t)\propto t^{\frac 13}.$

The dynamics of the perturbed region of non-zero curvature are described by
another Friedmann equation, with a scale-factor $S(\tau )$ 
\begin{eqnarray}
\left( \frac{dS}{d\tau }\right) ^2=\frac{8\pi G}3(\rho +\delta \rho )S^2-k,
\label{e3}
\end{eqnarray}
where $\tau $ is the proper time within the density perturbation, $\delta
\rho $, and the constant, $k$, measures its spatial curvature.

The proper time, $\tau ,$ inside the perturbation and that in the background
universe, $t,$ can be related using the equation of relativistic hydrostatic
equilibrium \cite{harr} \cite{ref3} 
\begin{eqnarray}
\frac{\partial \Phi ^{(grav)}}{\partial r}=-\frac{\partial p/\partial r}{%
p+\rho }  \label{hydrost}
\end{eqnarray}
where $\Phi ^{(grav)}$ is the (Newtonian) gravitational potential, and $r$
is the radial distance, and so

\begin{eqnarray}
d\tau =\exp [\Phi ^{(grav)}]dt.  \label{times}
\end{eqnarray}

The equation for hydrostatic equilibrium for a perfect fluid is derived
under the assumptions that the configuration is static (that is, only
spatial derivatives are non-vanishing) and that the gravitational field is
weak, so that the Newtonian gravitational potential $\Phi ^{(grav)}$
completely determines the metric. In this case equation (\ref{hydrost})
follows directly from the conservation of energy-momentum for a perfect
fluid. In this particular matching problem of two Friedmann universes
staticity means that there should be no radial flow of matter in or out of
the perturbation.

Equations (\ref{hydrost}) and (\ref{times}) imply 
\begin{eqnarray}
\frac{d\tau }{dt}=\left( \frac SR\right) ^3[1+\delta _0]^{-\frac 12},
\label{e4}
\end{eqnarray}
where we have introduced the constant $\delta ,$ the density contrast
parameter, defined by 
\[
\delta \equiv \frac{\delta \rho }\rho ,\;\;\;\;\;\;\;\;\;\;\;\;\mid \delta
\mid <1. 
\]
For the case of an overdensity we have $\delta >0$.

Assume that at some initial time $t_0$ the perturbation appears and the
following matching conditions hold between the metric scale factors inside
and outside the perturbation, \cite{steph}, 
\begin{eqnarray}
S_0=R_0\;\;\;\;\;\;\;\;\;\;\;\;\left( \frac{dS}{d\tau }\right) _0=\left( 
\frac{dR}{dt}\right) _0.  \label{mat}
\end{eqnarray}
Then, we have 
\begin{eqnarray}
\left( \frac{dR}{dt}\right) ^2 &=&\frac{\tilde G}{R^4}, \\
\left( \frac{dS}{d\tau }\right) ^2 &=&\frac{\tilde G}{S^4}(\beta -ES^4),
\label{e5}
\end{eqnarray}
where we have defined three new constants by 
\[
\tilde G\equiv \frac{8\pi G}3\rho _0R_0^6\;,\;\;\;\;\;\;\;\;\beta =1+\delta
_0\;,\;\;\;\;\;\;\;\;E\equiv \frac{\delta _0}{R_0^4}. 
\]

Furthermore, this implies that $S$ is given in terms of the background
scale-factor, $R,$ by 
\begin{eqnarray}
\frac{dS}{dR}=\frac SR(1-FS^4)^{\frac 12},  \label{e6}
\end{eqnarray}
where we have defined a further constant by 
\begin{eqnarray}
F\equiv \frac{\delta _0}{R_0^4(1+\delta _0)}.  \label{F}
\end{eqnarray}

Equation (\ref{e5}) implies that the maximum of the scale-factor of the
fluctuation is $S_{+}=F^{-\frac 14},$ and hence the integral of (\ref{e6})
is given by 
\begin{eqnarray}
S=S_{+}\frac{\sqrt{2}(R/R_{*})}{\sqrt{1+(R/R_{*})^4}}  \label{e6a}
\end{eqnarray}
where $R_{*}$ is the value of $R$ at $S_{+}$. The fluctuation thus begins
expanding with the background but is slowed with respect to it because of
its overdensity. Eventually, its expansion is halted by its self-gravity and
it begins to collapse whilst the background continues to expand.

We see that the spherical fluctuation has vanishing scale-factor $S$, i.e.
zero radius, for $R=0$ and also as $R\rightarrow \infty ,$ where $S\propto
R^{-1}$. This means that the perturbation does not collapse to a black hole
in finite proper time, since $R\propto t^{\frac 13}$. Physically, this
situation arises because the $p=\rho $ fluid possesses a Jeans length equal
to the horizon size and the fluid pressure is able to resist gravitational
collapse as soon as the fluctuation enters the horizon. This situation is
distinctive compared to the evolution of overdensities when $p<\rho .$ In
the $p<\rho $ case significant overdensities collapse to form black holes if
they turn-around when they are intermediate in scale between the particle
horizon and the Jeans scale \cite{carr}. In practice, even in the $p=\rho $
case some black-hole formation might be possible in finite time because of
small fluctuations in the sound speed and the horizon size but,
realistically, we would expect shock formation to play a role in damping
non-linear inhomogeneities in the $p<\rho $ cases (see also the discussions
of the scaling properties of black hole formation found by Choptuik and
others \cite{chop} in this connection).

Our description of the evolution of the spherical overdensity is completed
by deriving the ratio of the density in the perturbation to that in the
background universe. This is given by 
\begin{eqnarray}
\frac{\rho ^{(pert)}}{\rho ^{(back)}}=(1+\delta _0)\left( \frac RS\right) ^6.
\nonumber
\end{eqnarray}
Using equation (\ref{e6a}), this can also be expressed as 
\begin{eqnarray}
\frac{\rho ^{(pert)}}{\rho ^{(back)}}=(1+\delta _0)\left( \frac{R_{*}}{\sqrt{%
2}S_{+}}\right) ^6\left[ 1+\left( \frac R{R_{*}}\right) ^4\right] ^3.
\label{rat}
\end{eqnarray}

If we take the limit of small time (or small $R$) then we recover the usual
description of the growth of small perturbations in time in an appropriate
gauge.

Finally, it is interesting to note a simple duality scaling property that
appears in the above analysis. To see it in context, we can generalise the
analysis given above to the case of inhomogeneities in a general perfect
fluid model with equation of state $p=(\gamma -1)\rho .$ Equation (\ref{e6})
then generalises to 
\begin{eqnarray}
\left( \frac{dS}{dR}\right) ^2=\left( \frac SR\right) ^{3\gamma -4}(1+\delta
_0)^{(2-\gamma )/\gamma }[1-\left( \frac S{S_{+}}\right) ^{3\gamma -2}]
\label{e7}
\end{eqnarray}

We see that only in the case of a stiff perfect fluid, that is $\gamma =2$,
does (\ref{e7}) admit both a scaling symmetry ($S\rightarrow \alpha S$, $%
R\rightarrow \alpha R,\alpha $ constant) and the duality invariance 
\begin{eqnarray}
R\rightarrow R^{-1}.
\end{eqnarray}

In Figure 1 we show the evolution of $R,R^{-1},$ and $S$ with respect to the
background proper time, $t.$ The background expands as a flat FRW universe ($%
R\propto t^{1/3}$), whilst the overdensity expands less rapidly, reaches an
expansion maximum, but then collapses more slowly, tending to zero size at
an infinite future time.

\section{Connection with String Cosmology}

Since FRW universes are spherically symmetric about each point, the LCS
contributions to the antisymmetric tensor field strength vanish, and the
dilaton and axion fields behave as two coupled stiff ($\gamma =2$) perfect
fluids in the Einstein frame.  The Einstein frame is related to the string
frame by a conformal transformation of the metric of the form, 
\begin{eqnarray}
g_{\mu \nu }^{Einstein}=e^{-\Phi }g_{\mu \nu }^{String},  \label{conf}
\end{eqnarray}
where $\Phi $ is the dilaton. The advantage of the Einstein frame is that it
is the frame in which the action takes the Einstein-Hilbert form and so, if
there are no LCS contributions, the problem is that of general relativity
with an energy-momentum tensor consisting of two coupled stiff perfect
fluids. However, in order to provide a physical interpretation of the
dynamics, we should go back to the string frame. In string theory the
equation of motion for a classical test string implies that its world sheet
is a minimum surface with respect to the string metric; that is, the
(two-dimensional) analog of a geodesic for a point particle. Therefore, it
appears that strings ``see'' the string metric rather than the Einstein
metric \cite{ref4}.

The low-energy effective action in the Einstein frame yields to the
following set of equations ($\kappa ^2\equiv 8\pi G,c\equiv 1$) 
\begin{eqnarray}
R_{\mu \nu }-\frac 12g_{\mu \nu }R &=&\kappa ^2(^{(\Phi )}T_{\mu \nu
}+^{(H)}T_{\mu \nu })  \label{el1} \\
\nabla _\mu (e^{-2\Phi }H^{\mu \nu \lambda }) &=&0 \\
\Box \Phi +\frac 16e^{-2\Phi }H_{\mu \nu \lambda }H^{\mu \nu \lambda } &=&0
\label{el3}
\end{eqnarray}
where 
\begin{eqnarray}
^{(\Phi )}T_{\mu \nu } &=&\frac 12(\Phi _{,\mu }\Phi _{,\nu }-\frac 12g_{\mu
\nu }(\partial \Phi )^2) \\
^{(H)}T_{\mu \nu } &=&\frac 1{12}e^{-2\Phi }(3H_{\mu \lambda \kappa }H_\nu
^{\;\;\;\lambda \kappa }-\frac 12g_{\mu \nu }H_{\alpha \beta \gamma
}H^{\alpha \beta \gamma })
\end{eqnarray}

In four dimensions, if we take the space-time dual of the antisymmetric
tensor field strength $H_{\mu \nu \lambda },$ the Bianchi identity $dH=0$
shows that the dynamical content of the antisymmetric tensor field strength
is determined by a (pseudo-) scalar field, namely the axion field $b$. So,
we may write 
\[
H^{\mu \nu \lambda }=e^{2\Phi }\epsilon ^{\mu \nu \lambda \kappa }b_{,\kappa
} 
\]
where $\epsilon ^{\mu \nu \lambda \kappa }$ is the totally antisymmetric
Levi-Civita symbol. Furthermore, in the pure dilaton model we will discuss,
the perfect fluid is characterised by equal density and pressure, $p=\rho
=\frac 14\dot \Phi ^2$, and the 4-velocity obeys $u_\alpha u^\alpha =-1$.

Solutions for flat and closed FRW models have been discussed in \cite{ref5}.
Using conformal time $\eta $ the metric for a FRW model with a scale factor $%
a(\eta )$ can be written as 
\begin{eqnarray}
ds^2=a^2(\eta )(-d\eta ^2+\frac{dr^2}{1-kr^2}+r^2d\Omega ^2).
\end{eqnarray}

Naturally, the general-relativistic scenario discussed in the previous
section translates directly into a description of the post-big-bang era.
However, since the pre-big-bang model makes use of the invariance of
solutions under time reversal, it is misleading to make reference to $S_{+}$
and $R_0=S_0$ since these are two ``time-ordered'' length scales. If initial
conditions are specified at $R_0=S_0$ and developed into a state at $S_{+}$
then the time reversal causes $S_{+}$ to lie before $R_0=S_0;$ that is,
conditions at $R_0$ can no longer be treated as initial conditions.
Therefore, in the discussion of pre-big-bang solutions only the initial
scale $R_0=S_0$ should be used in the solutions for the scale factors.

\section{Post-Big-Bang Solutions}

For simplicity, only overdense perturbations $(\delta \rho >0,k>0)$ will be
discussed in this section. The symbol $\tau $ denotes conformal time in the
flat background universe while $\tau ^{(in)}$ is the conformal time inside
the spherical perturbation. Furthermore, the index ``*'' refers to the epoch
when the scale factor of the perturbation $S$ has its maximum value, $S_{+}$%
. Proper time runs from $0\rightarrow +\infty $. Solutions for a pure
dilaton model are given in \cite{ref5}.

(i) {\it Outside the perturbation, modelled as a flat FRW universe:}

The scale factor is given by 
\begin{eqnarray}
R^2(\tau )=R_{*}^2\frac \tau {\tau _{*}}
\end{eqnarray}
and the dilaton evolves as 
\begin{eqnarray}
e^\Phi =\left( \frac \tau {\tau _{*}}\right) ^{\pm \sqrt{3}}
\end{eqnarray}

(ii){\it \ Inside the perturbation, modelled as a closed FRW universe}

Introducing a function $\eta $ of the conformal time $\tau ^{(in)}$ inside
the perturbation defined by

\begin{eqnarray}
\eta (\tau ^{(in)})=\tan (\tau ^{(in)}),  \nonumber
\end{eqnarray}
the scale factor is given by 
\begin{eqnarray}
S^2(\eta )=2S_{+}^2\frac \eta {1+\eta ^2}
\end{eqnarray}
and the dilaton behaves as 
\begin{eqnarray}
e^\Phi =\eta ^{\pm \sqrt{3}},  \label{24}
\end{eqnarray}
since $\eta _{*}=1$.

In order to relate the conformal times inside and outside the fluctuation,
one can make use of equation (\ref{e6a}), which gives 
\begin{eqnarray}
\frac \tau {\tau _{*}}=\eta .
\end{eqnarray}
Furthermore, the evolution of the background scale-factor in proper time $t,$
in the Einstein frame, is found after integrating $dt=Rd\tau ,$ to be 
\begin{eqnarray}
R=R_{*}\left( \frac t{t_{*}}\right) ^{\frac 13}
\end{eqnarray}

As discussed above, the frame appropriate for physical interpretation is the
string frame. In this frame, the background scale factor is given by 
\begin{eqnarray}
^{(S)}R=e^{\Phi /2}R=R_{*}\left( \frac \tau {\tau _{*}}\right) ^{\frac
12(1\pm \sqrt{3})}.
\end{eqnarray}
Expressed in proper time in the string frame, i.e. after integrating $%
d^{(S)}t=\;^{(S)}Rd\tau $, $^{(S)}R(^{(S)}t)$ is given by

\begin{eqnarray}
^{(S)}R=R_{*}\left(\frac{^{(S)}t}{^{(S)}t_{*}} \right)^{\pm\frac{1}{\sqrt{3}}%
}.  \label{29}
\end{eqnarray}

Pre- and post-big-bang, respectively, refer to the background universe. In
the post-big-bang epoch the upper sign in (\ref{29}) is chosen. The matching
conditions (cf (\ref{mat})) provide a well-behaved metric in the Einstein
frame. Since physics should not be frame dependent, it is expected that
similar matching conditions should hold in the string frame. As one can
show, this implies that the sign chosen in (\ref{24}) for the evolution of
the dilaton inside the perturbation should be the same as that in the
background universe; hence

\begin{eqnarray}
\exp\left(\Phi^{(back)}\right)=\exp\left( \Phi^{(pert)}\right).  \label{rela}
\end{eqnarray}

The energy densities in the Einstein frame and the string frame can be
related by going back to the definition of the energy-momentum tensor as a
variational derivative of the Lagrangian ${\cal {L}}$ \cite{ref7}. In the
Einstein frame, we have 
\begin{eqnarray}
\rho =g^{00}T_{00}=(-g)^{-\frac 12}g^{00}\frac{\delta {\cal L}}{\delta g^{00}%
}.  \nonumber
\end{eqnarray}
Using the conformal transformation given by equation (\ref{conf}), we find
that the energy density in the string frame is given by 
\begin{eqnarray}
^{(S)}\rho =e^{-2\Phi }\rho .  \label{dens}
\end{eqnarray}
Hence, using (\ref{rela}), we confirm that the ratio of the energy density
in the perturbation to that in the background is the same in the Einstein
and the string frames: 
\begin{eqnarray}
\frac{^{(S)}\rho ^{(pert)}}{^{(S)}\rho ^{(back)}}=\frac{\rho ^{(pert)}}{\rho
^{(back)}}.
\end{eqnarray}

\section{Pre-Big-Bang Solutions}

In order to find solutions depending on one length scale we begin again with
equation (\ref{e6}). Furthermore, since the analysis does not depend on a
maximal length scale, the solutions for underdense regions (modelled by
matching to an open FRW universe, with $k<0$), can be treated as well.

\subsection{Underdense Regions}

In this case the density parameter $\delta $ is negative and hence the
constant $F$, defined in (\ref{F}), is also negative. Equation (\ref{e6}) 
now becomes 
\begin{eqnarray}
\frac{dS}{dR}=\frac SR\left( 1+\mid F\mid S^4\right) ^{\frac 12}.
\end{eqnarray}
This integrates to give 
\begin{eqnarray}
\frac S{S_0}=\frac{\sqrt{2}\left[ (1+\delta _0)^{-\frac 12}-1\right] ^{\frac
12}\left( \frac R{R_0}\right) }{\left[ \frac{\mid \delta _0\mid }{1+\delta _0%
}-\left[ (1+\delta _0)^{-\frac 12}-1\right] ^2\left( \frac R{R_0}\right)
^4\right] ^{\frac 12}}.  \label{un}
\end{eqnarray}

Hence, the ratio of the density in the fluctuation to that in the background
(in the Einstein frame) is given by 
\begin{eqnarray}
\frac{\rho ^{(pert)}}{\rho ^{(back)}}=\frac{1+\delta _0}{8\left[ (1+\delta
_0)^{-\frac 12}-1\right] ^3}\left[ \frac{\mid \delta _0\mid }{1+\delta _0}%
-\left[ (1+\delta _0)^{-\frac 12}-1\right] ^2\left( \frac R{R_0}\right)
^4\right] ^3.
\end{eqnarray}

\subsection{Overdense Regions}

In this case both the density parameter, $\delta ,$ and the constant $F$ are
positive. An integration of equation (\ref{e6}) yields 
\begin{eqnarray}
\frac S{S_0}=\frac{\sqrt{2}\left[ 1-(1+\delta _0)^{-\frac 12}\right] ^{\frac
12}\left( \frac R{R_0}\right) }{\left[ \frac{\delta _0}{1+\delta _0}+\left[
1-(1+\delta _0)^{-\frac 12}\right] ^2\left( \frac R{R_0}\right) ^4\right]
^{\frac 12}},  \label{ov}
\end{eqnarray}
and the ratio of the densities in the Einstein frame is given by 
\begin{eqnarray}
\frac{\rho ^{(pert)}}{\rho ^{(back)}}=\frac{1+\delta _0}{8\left[ 1-(1+\delta
_0)^{-\frac 12}\right] ^3}\left[ \frac{\delta _0}{1+\delta _0}+\left[
1-(1+\delta _0)^{-\frac 12}\right] ^2\left( \frac R{R_0}\right) ^4\right] ^3.
\end{eqnarray}

\vspace{2cm}

In order to connect the conformal times inside and outside the perturbation,
and hence the solutions for the pure dilaton model, we will use the original
notation of \cite{ref5}.

(i) {\it In the flat background} the scale factor is given by \cite{ref5}

\begin{eqnarray}
R^2=\frac K{\sqrt{3}}\tau
\end{eqnarray}
with $K$ some constant and $\tau $ the conformal time in the background. The
evolution of the dilaton is given by 
\begin{eqnarray}
e^\Phi =\left( \frac \tau {\tau _f}\right) ^{\pm \sqrt{3}}
\end{eqnarray}
where $\tau _f$ is an integration constant. For pre-big-bang solutions the
-- sign is chosen. In terms of proper time in the Einstein frame these
solutions read 
\begin{eqnarray}
e^\Phi =\left( \frac t{t_f}\right) ^{-\frac 2{\sqrt{3}}}
\end{eqnarray}
and 
\begin{eqnarray}
R^2=\left( \frac 32\frac K{\sqrt{3}}t\right) ^{\frac 23}.
\end{eqnarray}

Recalling that time is negative in the pre-big-bang era, one sees that in
the Einstein frame the background universe is contracting and the dilaton is
described by a growing function. The scale factor in the string frame is
given by 
\begin{eqnarray}
^{(S)}R^2=e^\Phi R^2=\left( \frac 32\right) ^{\frac 23}t_f^{\frac 2{\sqrt{3}%
}}\left( \frac K{\sqrt{3}}\right) ^{\frac 23}t^{\frac 23(1-\sqrt{3})},
\end{eqnarray}
and hence the universe is expanding in the string frame.

(ii) {\it Inside the density fluctuation} the scale factor is given by

\begin{eqnarray}
S^2=\frac{\tilde K}{\sqrt{3}}\frac \eta {(1+k\eta ^2)}^{},
\end{eqnarray}
with $\tilde K$ some constant, and 
\[
\eta =\left\{ 
\begin{array}{llr}
\tan (\tau _{in}), & k=+1 & {\rm overdensity} \\ 
\tanh (\tau _{in}), & k=-1 & {\rm underdensity}
\end{array}
\right. 
\]
where $\tau _{in}$ is the conformal time inside the fluctuation. Using the
matching conditions, (\ref{mat}), $\tau _0$ and $\eta _0$ can be related at
this epoch by 
\begin{eqnarray}
\tau _0=\frac{\eta _0}{1-k\eta _0^2}.  \label{tau}
\end{eqnarray}

The solution for the dilaton is found to be 
\begin{eqnarray}
e^\Phi =\left( \frac \eta {\eta _f}\right) ^{-\sqrt{3}},  \label{dil}
\end{eqnarray}
with $\eta _f$ an integration constant.

Since the dilaton solutions are involved in the transformation of the energy
densities from the Einstein frame to the string frame, it is enough to
relate $\tau $ and $\eta $ (and not explicitly the conformal times inside
and outside the fluctuation) and this can be done using equations (\ref{un})
and ({\ref{ov}). }First, note that 
\begin{eqnarray}
\left( \frac R{R_0}\right) ^2=\frac \tau {\tau _0}  \label{R}
\end{eqnarray}
and 
\begin{eqnarray}
\left( \frac S{S_0}\right) ^2=\frac{\eta (1+k\eta _0^2)}{\eta _0(1+k\eta ^2)}%
\ .  \label{S}
\end{eqnarray}

Observing that $\delta _0$ is positive or negative, according to the sign of 
$k,$ equations (\ref{un}) and (\ref{ov}) can be reduced to a single
expression, namely 
\begin{eqnarray}
\left( \frac S{S_0}\right) ^2=\frac{2\alpha \left( \frac R{R_0}\right) ^2}{%
\gamma +\alpha ^2\left( \frac R{R_0}\right) ^4}\equiv \Gamma (\frac \tau
{\tau _0})
\end{eqnarray}
with 
\[
\alpha \equiv 1-(1+\delta _0)^{-\frac 12}\;\;\;,\;\;\;\;\;\;\gamma \equiv 
\frac{\delta _0}{1+\delta _0}.
\]
The function $\Gamma (\frac \tau {\tau _0})$ is defined in accord with
equation (\ref{R}). So, using equation (\ref{S}), the function $\eta $ can
be expressed in terms of the conformal time $\tau $ in the background
universe, 
\begin{eqnarray}
\eta =k\frac{1-\sqrt{1-4kE_0^2\Gamma ^2}}{2E_0\Gamma },
\end{eqnarray}
where 
\[
E_0\equiv \frac{\eta _0}{1+k\eta _0^2}
\]
which can be expressed in terms of $\tau _0$ using (\ref{tau}).

Therefore, in this more general case, the relationship between the dilaton
solutions inside and outside the fluctuation is more complicated than that
found earlier for the post-big-bang era (cf equation (\ref{rela})).

The perturbation originates at some epoch $\tau _0$ with a scale factor
corresponding to a value $R_0$ in the background universe. The big-bang
occurs at $\tau =0$, but one should not follow the evolution all the way to $%
\tau =0$ since at some point the string coupling becomes too strong for the
low-energy action to remain a valid approximation to the full theory. In the
pre-big-bang phase the proper time is restricted to $t<0$. Suppose that a
perturbation originates at $t_0$ and the period of interest is between $t_0$
and some time $t_s,$ (with $\mid t_s\mid <\mid t_0\mid $), when the string
coupling becomes strong, so we are interested in the regime $\tau <\tau _0$, 
$R<R_0$. Assuming $R\ll R_0,$ the scale factor of the perturbation $S$ and
the function of its conformal time $\eta (\tau )$ can be expanded as
follows, 
\begin{eqnarray}
\frac S{S_0} &\simeq &\left( \frac{2\alpha }\gamma \right) ^{\frac 12}\frac
R{R_0}  \label{S/S0} \\
\frac \eta {\eta _0} &\simeq &\frac{2\alpha }\gamma (1+k\eta _0^2)^{-1}\frac
\tau {\tau _0}.  \label{e/e0}
\end{eqnarray}

Thus, $\frac S{S_0}\sim \frac R{R_0}$, $\frac \eta {\eta _0}\sim \frac \tau
{\tau _0}$. However, for $\delta _0>0$, the proportionality factor in (\ref
{S/S0}) is bigger then unity, and so the scale factor of the perturbation
exceeds that of the background universe (see figure 2).

This might be interpreted as interchanging the role of the flat and the
closed FRW models ({\it ie} there are inhomogeneities corresponding to flat
space-time sections in a closed FRW background universe \cite{btip}).
Figures 3 and 4 show the behaviour of the two scale factors, $R$ and $S,$
with time, $t$, for different choices of the initial density parameter $%
\delta _0$. In Figure 3, the scale factor of the overdense region is larger
than that of the background universe within the period of interest between $%
t_0$ up to some time, $t_s$. In Figure 4, where there is an initial
underdensity at $t_0,$ the scale factor of the perturbation stays below that
of the background universe.

It is clear that the evolution of the two scale factors is very similar,
from the origin of the perturbation up to the big bang. Thus, one cannot
actually say that the perturbation evolves independently from the
background. Qualitatively, the same behaviour is found in the string frame.
The evolution of the dilaton depends on the respective conformal times (or
functions of them), namely $\tau $ and $\eta $. However, in the regime
between the origin of the perturbation and the big bang these are directly
proportional to each other as can be seen from (\ref{e/e0}). Hence, the
evolution of the scale factors in the string frame is very similar as well.

In accord with the behaviour in the general relativistic (hence post-big
bang) regime, no primordial black holes should be formed in the pre-big-bang
era in this model since the perturbation does not evolve independently and
so does not form trapped surfaces during the collapse of the scale factor of
the background universe in the Einstein frame.

From this discussion it can be concluded that the introduction of a
spherical perturbation modelled as a closed or open FRW model in a flat FRW
model does not destroy the global isotropy of the flat background universe.
Hence the pre-big-bang flat FRW solution is robust with respect to this
special type of curvature perturbation.

In heterotic string theory, any solutions containing only a dilaton can be
transformed into solutions containing a dilaton and axion without changing
the background metric. S-duality reflects the fact that heterotic string
theory is invariant under $SL(2,I\!R)$ transformations. Define a complex
scalar $\lambda =b+ie^{-\Phi }$ then \cite{ref6} 
\[
\lambda \rightarrow \lambda ^{\prime }=\frac{\alpha \lambda +\beta }{\gamma
\lambda +\delta }\;\;\;\;\;\;\;\;g_{\mu \nu }\rightarrow g_{\mu \nu
}\;\;\;\;\;\;\;\;\;\;\alpha ,\beta ,\gamma ,\delta \in I\!R,\;\;\alpha
\delta -\beta \gamma =1 
\]
is a solution to the low-energy equations of motion (equations (\ref{el1})-(%
\ref{el3})).

If we start with a pure dilaton solution 
\[
\lambda =ie^{-\Phi } 
\]
and choose $\alpha =\beta =\gamma =-\delta =1/\sqrt{2}$, then $\lambda
^{\prime }=b^{new}+i\exp [-\Phi ^{new}]$ is given by 
\begin{eqnarray}
b^{new} &=&\tanh \Phi \\
e^{\Phi ^{new}} &=&\cosh \Phi .
\end{eqnarray}
Hence, the pure FRW dilaton solution has 
\[
e^\Phi =\left( \frac \tau {\tau _0}\right) ^{\pm \sqrt{3}}, 
\]
where $\tau $ is a different function of the conformal time $\eta $ for each
of the FRW models which can be read of from the solutions given above, and
is related to a FRW dilaton-axion solution with 
\begin{eqnarray}
e^\Phi &=&\frac 12\left[ \left( \frac \tau {\tau _0}\right) ^{\sqrt{3}%
}+\left( \frac \tau {\tau _0}\right) ^{-\sqrt{3}}\right] , \\
b &=&\pm \frac{\left( \frac \tau {\tau _0}\right) ^{\sqrt{3}}-\left( \frac
\tau {\tau _0}\right) ^{-\sqrt{3}}}{\left( \frac \tau {\tau _0}\right) ^{%
\sqrt{3}}+\left( \frac \tau {\tau _0}\right) ^{-\sqrt{3}}},
\end{eqnarray}
which is given in \cite{ref5}.

However, near the big-bang (that is, for small $\frac \tau {\tau _0}),$ the
dilaton solution goes over into that obtained for the pure dilaton case and
so the conclusions from the pure dilaton case are not likely to be changed
fundamentally.

\section{Conclusions}

A very simple inhomogeneous string cosmological model has been investigated
in the pre- and post-big-bang eras. In the post-big-bang era the usual
general relativistic behaviours of a growing overdensity or underdensity are
found. This leads to an increasingly inhomogeneous universe if the initial
inhomogeneities are significant, or they are not inflated away. However, in
the pre-big-bang phase the global isotropy of the background FRW model is
unaffected by the introduction of a spherically symmetric curvature
perturbation. This rather different behaviour of the model in the pre- and
post-big-bang stages is a manifestation of the fact that these two regimes
pick out different parts of the general relativistic solution. In general
relativity, considering only positive (proper) times, the solution is
naturally divided by the time $t_0$ when the perturbation originates. The
physical solution is then given for times later than $t_0$, whereas the part
between the big bang and $t_0$ is discarded. However, in the pre-big-bang
stage one is considering a time-reflection of the general relativistic
solution. Hence, the part of the solution that was discarded in the general
relativistic case (post-big-bang) becomes the physical solution in the
pre-big-bang era. Thus, it is the physical solution of general relativity
which is considered unphysical in the pre-big-bang regime. Hence, unlike in
the general relativistic solution, a curvature perturbation like that
considered here does not lead to an inhomogeneization of the universe in the
pre-big-bang epoch.

The authors of references \cite{venez} and \cite{ref3b} discussed
inhomogeneous pre-big-bang models using standard approximation techniques of
general relativistic cosmology. They found that the approximation of
neglecting spatial gradients becomes better as the big-bang singularity is
approached. This conclusion is confirmed by the simple quasi-homogeneous
model discussed here. The evolution of a curvature perturbation modelled by
a section of a Friedmann universe is very similar to that of the background
Friedmann universe and indeed becomes more similar to it with time. Hence
isotropy is conserved and spatial gradients do not become important. We note
also that part of the general solution for general relativistic cosmological
models containing a $p=\rho $ perfect fluid is a perturbation of a
Kasnerlike solution which (unlike for case of perfect fluids with equation
of state $p<\rho $) contains the isotropic expansion as a particular case.
Because of this, the general behaviour of the equations of string cosmology
at early times is significantly simpler than that of general relativity in
vacuum.

{\bf Acknowledgements}

JDB is supported by a PPARC Senior Fellowship. KEK is supported by the
German National Scholarship Foundation. We would like to thank Ed Copeland
and David Wands for helpful discussions.

\begin{figure}
\centerline{\psfig{file=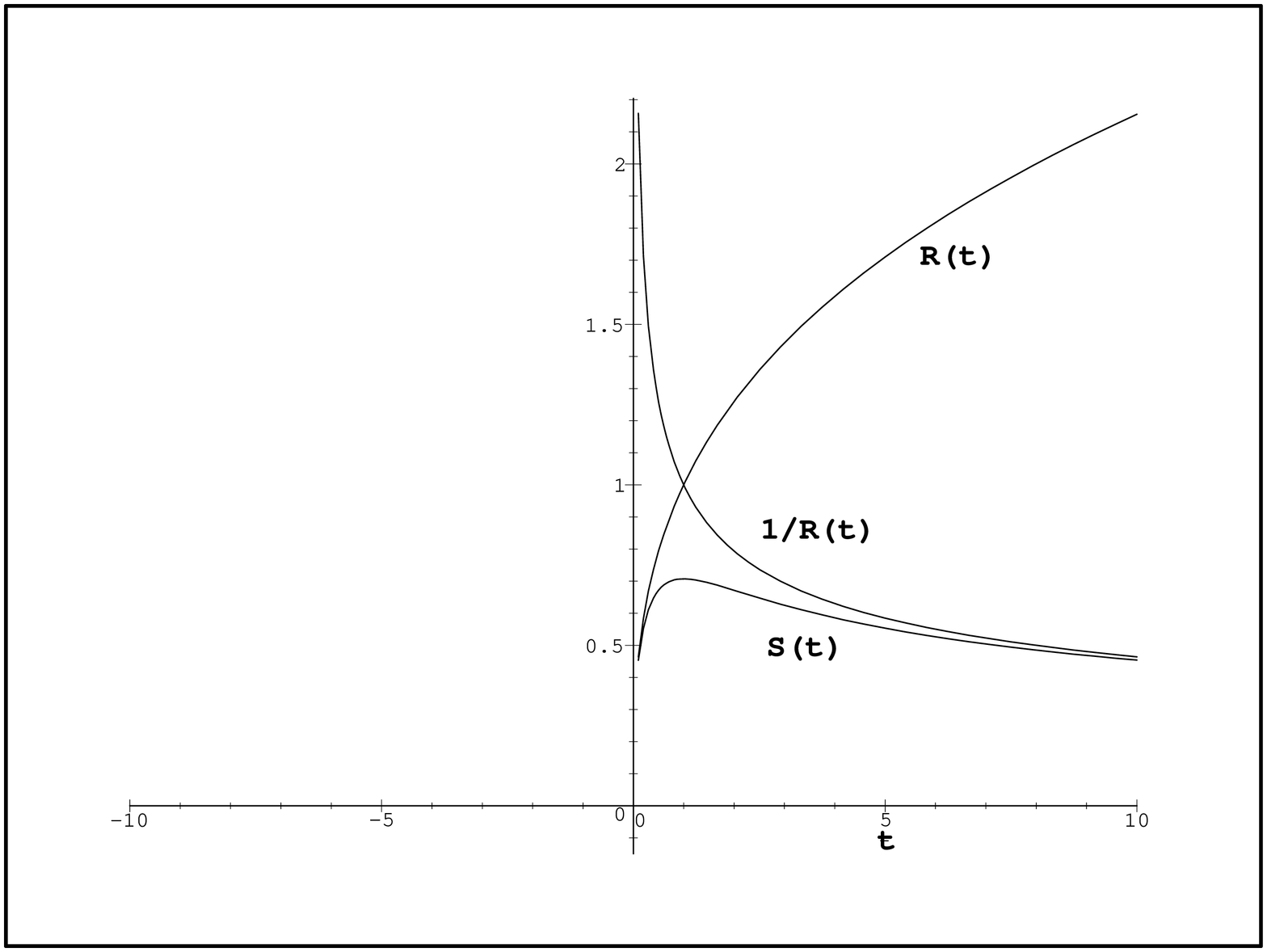,height=3. in,width=3. in,angle=-90}}
\caption{ Evolution of the background scale factor
$R\propto t^{\frac{1}{3}}$
, the transformed one $R^{-1}\propto t^{-\frac{1}{3}}$ and the scale factor
of the perturbation $S(t)\propto t^{\frac{1}{3}}
(1+t^{\frac{4}{3}})^{-\frac{1}{2}}$ in terms of proper background time $t$. }
\end{figure}

\begin{figure}
\centerline{\psfig{file=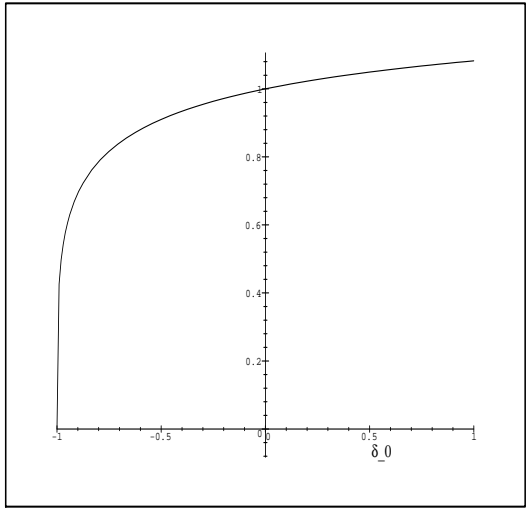,height=3. in,width=3. in,angle=-90}}
\caption{The proportionality factor $\left( 2\frac \alpha \gamma
\right) ^{\frac 12}$ as a function of $\delta _0$.}
\end{figure}

\begin{figure}
\centerline{\psfig{file=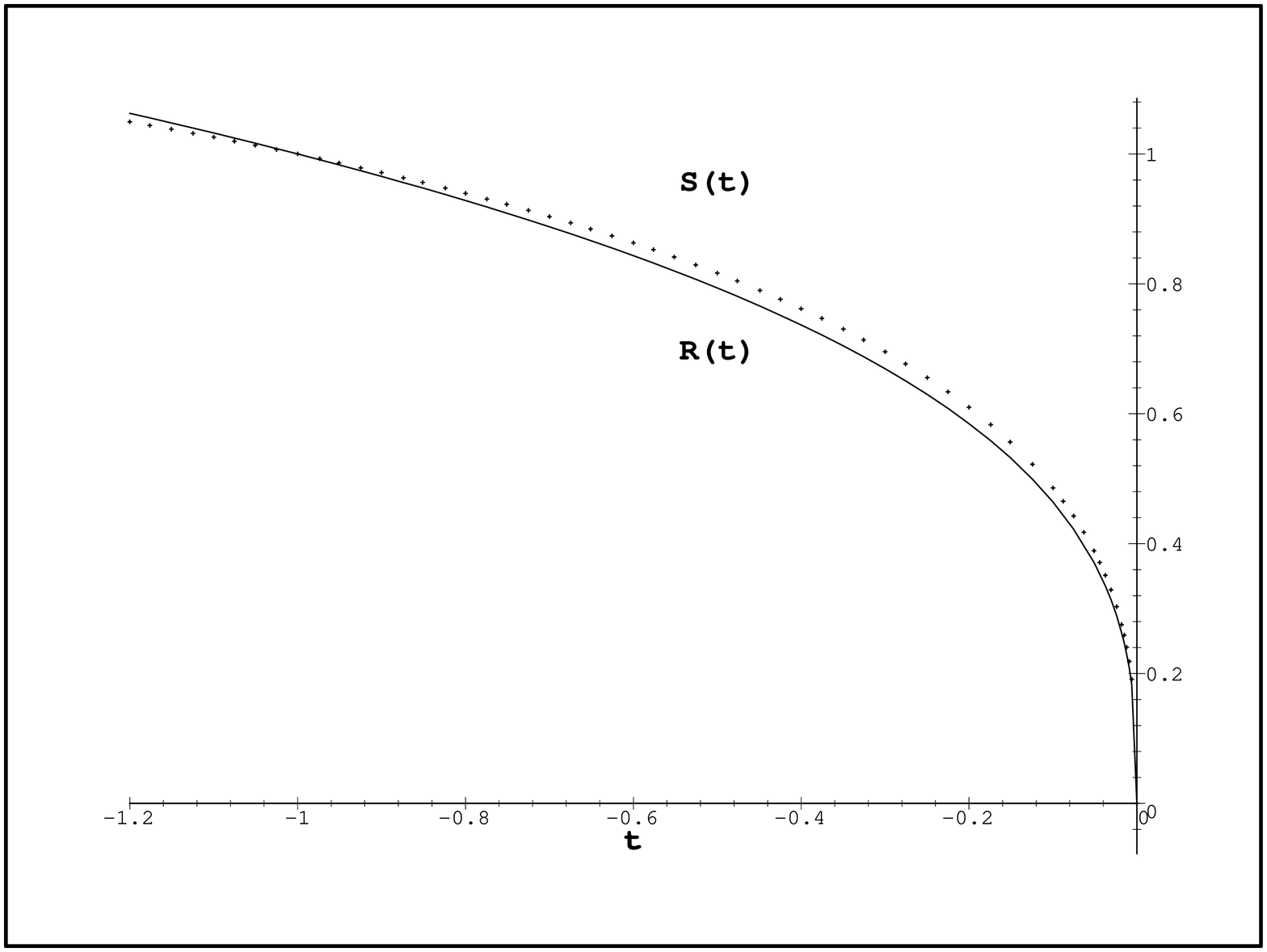,height=3. in,width=3. in,angle=-90}} 
\caption{Evolution of the scale factors of the perturbation $S$
and the background universe $R$ as a function of proper time $t$ in
the Einstein frame ;
$\delta _0=0.5$, $R_0=1$, $t_0=-1$.}
\end{figure}
 
\begin{figure}
\centerline{\psfig{file=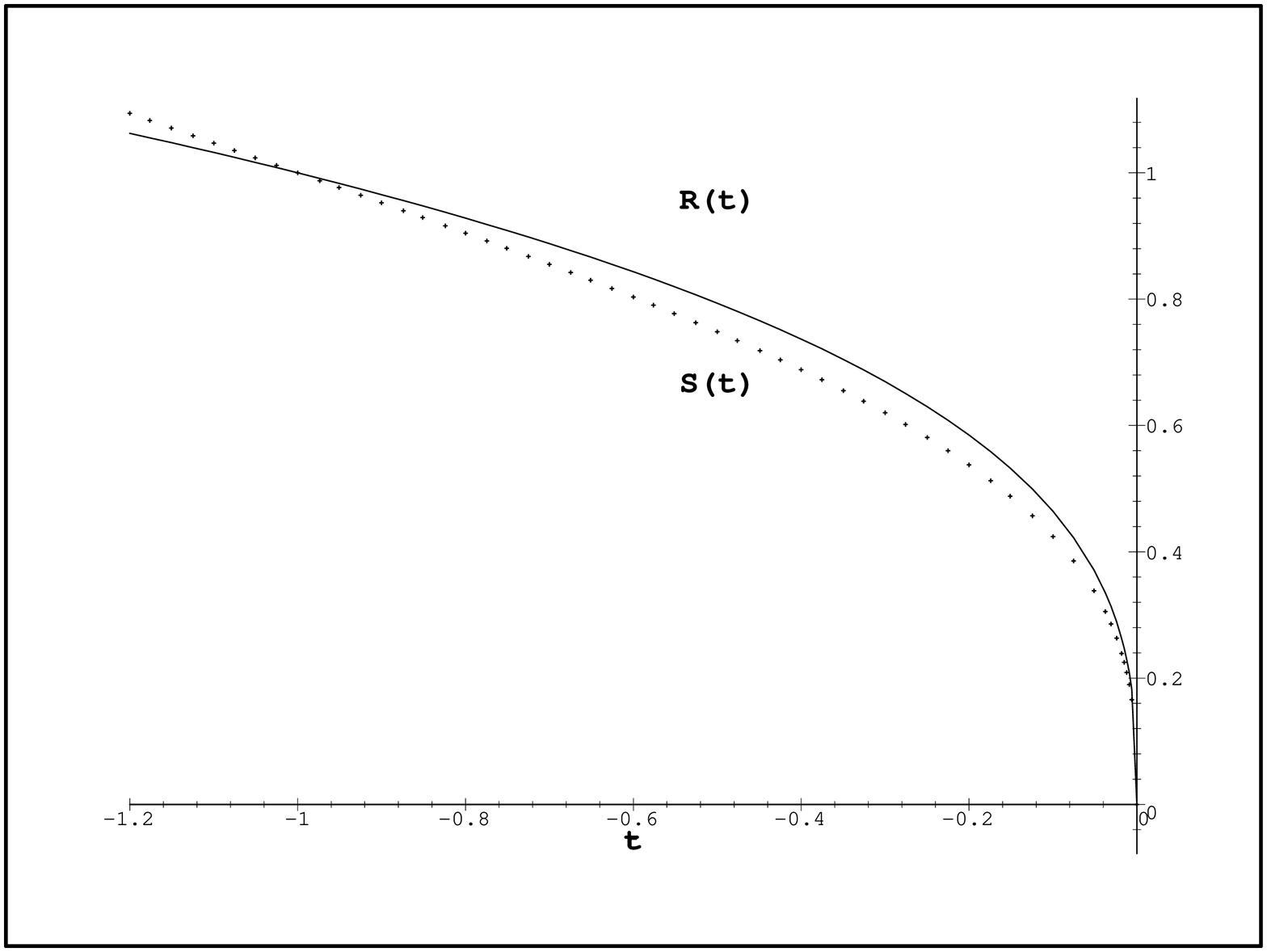,height=3. in,width=3. in,angle=-90}}
\caption{Evolution of the scale factors of the perturbation $S$ and
the background universe $R$ as a function of proper time $t$
in the Einstein frame; $%
\delta_{0}=-0.5$, $R_{0}=1$, $t_{0}=-1$.}
\end{figure}


\begin{thebibliography}{99}
\bibitem{refA}  An archive for string cosmology maintained by M. Gasperini
can be found at {\tt http://www.to.infn.it/teorici/gasperini/}

\bibitem{ref5}  E. J. Copeland, A. Lahiri, D. Wands, Phys. Rev. D {\bf 50,}
4868 (1994).

\bibitem{batak}  M. Gasperini, R. Ricci, Class. Quant. Grav. {\bf 12, }677
(1995); N.A. Batakis, Phys. Lett. B {\bf 353, }39 (1995); N.A. Batakis, A.A
Kehagias, Nucl. Phys. B {\bf 449, }248 (1995).

\bibitem{ref10}  J.D. Barrow and M. Dabrowski, Phys. Rev. D {\bf 55, }630
(1997).

\bibitem{ref11}  J.D. Barrow and K.E. Kunze, Phys. Rev. D {\bf 55, }623
(1997).

\bibitem{ref9}  J.D. Barrow and K.E. Kunze, Phys. Rev. D {\bf 56, }741
(1997).

\bibitem{ref8}  A. Feinstein, R. Lazkoz, and M.A. V\'azquez-Mozo, Phys. Rev.
D hep-th/9704173.

\bibitem{venez}  G. Veneziano, ``Inhomogeneous Pre-Big Bang String
Cosmology'' hep-th/9703150, Phys. Lett. B

\bibitem{ref3b}  A. Buonanno, K. A. Meissner, C. Ungarelli, G. Veneziano
``Classical Inhomogeneities in String Cosmology'' hep-th/9706221.

\bibitem{veldom}  The 'velocity-dominated' approximation, originally
introduced in a formal way by D. Eardley, R. Sachs and E.P.T. Liang, J.
Math. Phys. {\bf 13}, 99 (1972) and E.P.T. Liang, J. Math. Phys. {\bf 13,}
386 (1972), following its use by Landau and Lifshitz in {\it The Classical
Theory of Fields, }Pergamon, Oxford (1962) and E.M. Lifshitz and I.
Khalatnikov, Adv. Phys. {\bf 12}, 208 (1963). It was also introduced by K.
Tomita, Prog. Theo. Phys. {\bf 48}, 1503 (1972) under the name
'anti-Newtonian approximation', so called because neglecting space
derivatives ($\partial /\partial x$) with respect to time derivatives ($%
\partial /\partial (ct))\ $is equivalent to taking a limit in which the
speed of light goes to zero ($c\rightarrow 0$) --- the antithesis of
Newtonian physics in which $c=\infty .$

\bibitem{lemait}  G. Lema\^itre, Ann. Soc. Sci. Bruxelles A {\bf 53, }51
(1933).

\bibitem{harr}  E.R. Harrison Phys. Rev. D {\bf 1, }2726 (1970); E.R.
Harrison in {\it Carg\`ese Lectures in Physics, }Vol. 6, ed E. Schatzmann,
Gordon \& Breach, NY, (1973).

\bibitem{ref1}  M. B. Green, and J. H. Schwarz, Phys. Lett. {\bf 149B} 117
(1984); M. B. Green, J. H. Schwarz, and E. Witten {\sl Superstring Theory}
Volume 2 CUP (1987).

\bibitem{ref2}  B. A. Campbell, M. J. Duncan, N. Kaloper, and K. A. Olive,
Nucl. Phys. B {\bf 351} 778 (1991).

\bibitem{ref1a}  G. Veneziano, Phys. Lett. B {\bf 265} 287 (1991); M.
Gasperini, and G. Veneziano, Astropart. Phys. {\bf 1} 317 (1993); see also 
{\tt http://www.to.infn.it/teorici/gasperini/}

\bibitem{ref2a}  M. Gasperini, M. Maggiore, G. Veneziano ``Towards a
non-singular pre-big-bang cosmology'' hep-th/9611039; R. Brustein, R. Madden
``Graceful exit and energy conditions in string cosmology'' hep-th/9702043.

\bibitem{quant}  M. Gasperini, G. Veneziano, Gen. Rel. Grav. {\bf 28, }1301
(1996); M. Dabrowski, C. Kiefer, Phys. Lett. B {\bf 397, }185 (1997).

\bibitem{ref3}  L. D. Landau, and E. M. Lifshitz {\sl Classical Theory of
Fields, 4th edn.,} Pergamon, Oxford (1975).

\bibitem{steph}  For a summary of formal matching conditions in general
relativity see H. Stephani, {\it General Relativity}, CUP, Cambridge,
(1982), section 16.5.

\bibitem{carr}  B.J. Carr and S.W. Hawking, MNRAS {\bf 168}, 399 (1974),
B.J. Carr, Astrophys. J. {\bf 201}, 1 (1975).

\bibitem{chop}  M.W. Choptuik, Phys. Rev. Lett. {\bf 70,} 9 (1993); J.C.
Niemeyer and K. Jedamzik, astro-ph/9709072.

\bibitem{ref4}  J. A. Harvey, A. Strominger ``Quantum Aspects of Black
Holes'' 1992 Trieste Spring School on String Theory and Quantum Gravity;
1992 TASI Summer School in Boulder, Colorado; hep-th/9209055

\bibitem{ref7}  S. Kalara, N. Kaloper, and K. A. Olive Nucl. Phys. B {\bf 341%
} 252 (1990).

\bibitem{btip}  The global behaviour of such inhomogeneities and their
relation to the problem of when closed universes (ie those with compact
spatial hypersurfaces) recollapse is discussed by Y. B. Zeldovich and L.
Grishchuk, MNRAS {\bf 207, }23P (1984), J.D. Barrow and F.J. Tipler, MNRAS 
{\bf 216}, 395 (1985) and J.D. Barrow, G. Galloway, and F.J. Tipler, MNRAS 
{\bf 223}, 835 (1986).

\bibitem{ref6}  A. Sen, Int. J. Mod. Phys. A{\bf 9} 3707 (1994).
\end{thebibliography}
\end{document}